\documentclass[aps,12pt,onecolumn,preprintnumbers,amsmath,amssymb,superscriptaddress]{revtex4}

\usepackage[english]{babel}
\usepackage{graphicx}
\usepackage{epstopdf}
\usepackage{color}
\usepackage{ragged2e}
\usepackage{float}
\usepackage{xcolor}
\usepackage{times}
\usepackage{bm}
\usepackage[final]{pdfpages}
\begin{document}

\newcommand{\unit}[1]{\:\mathrm{#1}}            
\newcommand{\To}{\mathrm{T_0}}
\newcommand{\Tp}{\mathrm{T_+}}
\newcommand{\Tm}{\mathrm{T_-}}
\newcommand{\EST}{E_{\mathrm{ST}}}
\newcommand{\Rp}{\mathrm{R_{+}}}
\newcommand{\Rm}{\mathrm{R_{-}}}
\newcommand{\Rpp}{\mathrm{R_{++}}}
\newcommand{\Rmm}{\mathrm{R_{--}}}
\newcommand{\ddensity}[2]{\rho_{#1\,#2,#1\,#2}} 
\newcommand{\ket}[1]{\left| #1 \right>} 
\newcommand{\bra}[1]{\left< #1 \right|} 




\title{Quadrupolar excitons in a tunnel-coupled van der Waals heterotrilayer}

\author{Weijie Li*}
\affiliation{Department of Physics, Emory University, 30322 Atlanta, Georgia, USA}
\author{Zach Hadjri*}
\affiliation{Department of Physics, Emory University, 30322 Atlanta, Georgia, USA}
\author{Jin Zhang*}
\affiliation{Center for Free Electron Laser Science, Max Planck Institute for the Structure and Dynamics of Matter, 22761 Hamburg, Germany}
\author{Luka M. Devenica*}
\affiliation{Department of Physics, Emory University, 30322 Atlanta, Georgia, USA}
\author{Song Liu}
\affiliation{Department of Mechanical Engineering, Columbia University, 10027 New York, New York, USA}
\author{James Hone}
\affiliation{Department of Mechanical Engineering, Columbia University, 10027 New York, New York, USA}
\author{Kenji Watanabe}
\affiliation{Research Center for Functional Materials, National Institute for Materials Science, 1-1 Namiki, Tsukuba 305-0044, Japan}
\author{Takashi Taniguchi}
\affiliation{International Center for Materials Nanoarchitectonics, National Institute for Materials Science,  1-1 Namiki, Tsukuba 305-0044, Japan}
\author{Angel Rubio}
\affiliation{Center for Free Electron Laser Science, Max Planck Institute for the Structure and Dynamics of Matter, 22761 Hamburg, Germany}
\affiliation{Center for Computational Quantum Physics, Simons Foundation Flatiron Institute, 10010 New York, New York, USA}
\affiliation{Nano-BioSpectroscopy Group, Departamento de Fisica de Materiales,Universidad del Pa{\'i}s Vasco, 20018 San Sebastián, Spain}
\author{Ajit Srivastava$^{\dagger}$}
\affiliation{Department of Physics, Emory University, 30322 Atlanta, Georgia, USA}

\maketitle
$^{\dagger}$Correspondence to: ajit.srivastava@emory.edu
\\
$^{\ast}$These authors contributed equally to this work.


{\bf Strongly bound excitons~\cite{wangRMP2018colloquium,MakNMat2013, HePRL2014} and many-body interactions~\cite{RiveraScience2016,liNNano2021optical, kremserNpj2D2020discrete,sunNPhot2022excitonic} between them determine light-matter interactions in van der Waals (vdW) heterostructures of 2D semiconductors~\cite{FangPNAS2014,RiveraNComm2015, RiveraScience2016, WangNanoLett2018,CiarrocchiNP2019,BaranowskiNanoLett2017}. Unlike fundamental particles, quasiparticles in condensed matter, such as excitons, can be tailored to alter their interactions and realize emergent quantum phases. Here, using a WS$_2$/WSe$_2$/WS$_2$ heterotrilayer, we create a quantum superposition of oppositely oriented dipolar excitons -- a quadrupolar exciton -- wherein an electron is layer-hybridized in WS$_2$ layers while the hole localizes in WSe$_2$. In contrast to dipolar excitons~\cite{RiveraNComm2015}, symmetric quadrupolar excitons only redshift in an out-of-plane electric field, consistent with ab initio calculations, regaining dipolar characteristics at higher fields. Electric field tunes the hybridization and allows for lifetime control through modification of the excitonic wavefunction. Lack of density-dependent blue shift of heterotrilayer excitons compared to dipolar excitons is consistent with quadrupolar interactions. Our results present vdW heterotrilayers as a field-tunable platform to engineer light-matter interactions and explore quantum phase transitions between spontaneously ordered many-exciton phases~\cite{SlobodkinPRL2020,SammonPRB2019, astrakharchikPRB2021quantum}}. 



An exciton, which comprises of a Coulomb-correlated electron-hole pair, is an elementary excitation of a semiconductor resembling a hydrogen atom albeit with possible modifications arising from  lattice effects~\cite{yongNMat2019valley}. The resulting atom-like anharmonic spectra of excitons can be exploited for effective photon-photon interactions and nonlinearities mediated by light-matter coupling~\cite{guNComm2021enhanced,ZhangNature2021,TanPRX2020}. 
In addition to atom-like neutral excitons, excitonic complexes such as trions and biexcitons, resembling ions and molecules, add further richness to exciton physics~\cite{wangRMP2018colloquium}. Going beyond independent excitons, an open question in the spirit of condensed matter physics is -- In what quantum phases does a system of interacting excitonic complexes organize itself? Besides being of fundamental importance to understand phases of matter in a driven-dissipative setting~\cite{dagvadorjPRB2021first}, addressing this question can also enable generation of novel states of light. To this end, engineering excitonic structures with an aim to tune light-matter interactions and exciton-exciton interactions is crucial.


In vdW heterostructures of 2D semiconductors, the layer degree of freedom can endow excitons with additional internal structure, such as a static dipole moment in interlayer excitons of heterobilayers with type-II band alignment, affecting their interactions and light-matter coupling~\cite{RiveraNComm2015,ZhangNature2021}. However, this dipole moment in heterobilayers is fixed by sample geometry. In this study, by exploiting quantum tunneling of electrons to modify the excitonic structure, we realize quadrupolar excitons in vdW heterotrilayers. Moreover, an electric field is shown to continuously tune between quadrupolar and dipolar excitons. The field induced changes to excitonic wavefunction are manifested as modifications in excitonic lifetime and interactions.

Figure 1a shows the schematic of our WS$_2$/WSe$_2$/WS$_2$ heterotrilayer. If the outer layers are uncoupled, this heterostructure can be expected to host oppositely oriented dipolar interlayer excitons because of type-II band alignment between WSe$_2$ and WS$_2$. However, in the presence of a finite tunnel coupling, the electron is layer-hybridized between WS$_2$ layers, whereas the hole resides in the middle WSe$_2$ layer. The strength of tunneling $t$ determines the energy difference between the lower energy symmetric (Fig. 1b) and higher energy antisymmetric (Fig. 1c) hybridized electronic states. More importantly, the symmetric (antisymmetric) electronic state has a finite (vanishing) weight in the WSe$_2$ layer. Together with the hole in the middle layer, the resulting excitonic states have no net dipole moment but a quadrupole moment. In other words, electron tunneling hybridizes the two degenerate, oppositely oriented dipolar excitons into symmetric and antisymmetric quadrupolar excitons. However, quadrupolar excitons in heterotrilayers should be contrasted with earlier reports on hybridized interlayer and intralayer excitons~ \cite{AlexeevNature2019, ShimazakiNature2020, HsuSciAdv2019tailor} which do not have a quadrupolar moment. 

Figure 1d shows a microscope image of our heterotrilayer sample with dual gates for independent control of displacement field and carrier doping (see Methods). The heterostructure also has a region of heterobilayer which serves to directly compare dipolar and quadrupolar excitons. The stacking order is chosen to be AA for the outer WS$_2$ layers because AB stacking has negligible hybridization with spin-valley conserving tunneling~\cite{XiaoPRL2012}. Figure 1e-f shows low temperature ($\sim$ 6K) photoluminescence (PL) spectra of the trilayer and bilayer regions together with the corresponding photoluminescence excitation (PLE) spectra. Both the emission energy and excitation resonance energy are redshifted in the trilayer region compared to the bilayer region. The trilayer spectra are dominated by three peaks which we assign to, in order of decreasing energy, a spin-triplet IX, a spin-singlet IX and the spin-singlet IX phonon replica, based on earlier reports in WSe$_2$/WS$_2$ heterobilayer ~\cite{yuOptica2020observation, ParadisanosNComm2021}.

To distinguish between quadrupolar and dipolar excitons, we study the PL from bilayer and trilayer regions as a function of an out-of-plane electric field ($\mathbf{E} = E_{\mathrm{z}} \hat{\mathrm{z}}$). We emphasize that the bilayer and the trilayer region are under the same set of gates and hence are subject to nominally the same $\mathbf{E}$. As shown in Fig.~2a, the PL from the bilayer region shifts linearly, displaying both red and blueshifts depending on the direction of ($E_z$). Thus, the bilayer emission behaves as expected for dipolar excitons with an energy shift $\delta \mathcal{E} = - \mathbf{d} \cdot \mathbf{E}$~\cite{JaureguiScience2019}. Moreover, the slope of the energy shift, $d\mathcal{E}/d E_{\mathrm{z}}$, which is proportional to the dipole moment, is consistent with the layer ordering in the bilayer region. In stark contrast, the PL from trilayer region  redshifts for either direction of $\mathbf{E}$, with a characteristic nonlinear shape (Fig.~2b). While we measure $\mathbf{E}$-dependence of the PL with opposite voltages applied to the top and bottom gates so as to not introduce carriers in the sample, imperfections in gate configurations can lead to a small amount of unintentional doping which could also result in energy shifts. By monitoring the reflectance of intralayer excitonic resonances, which remain unchanged with $\mathbf{E}$, we conclude that accidental doping, if any, is not sizable to cause observed energy shifts of tens of meV (see Supplementary). However, we note that the PL in both the bilayer and trilayer regions for $V_{\mathrm{tg}}$ = $-V_{\mathrm{bg}} <$  -3V displays slight broadening and reduction in intensity compared to $V_{\mathrm{tg}}$ = $-V_{\mathrm{bg}} >$ 3V. We attribute this to light unintentional doping for $V_{\mathrm{tg}}$ = $-V_{\mathrm{bg}} <$ -3V, which causes a slight asymmetry in the nonlinear redshift about $\mathbf{E}$ = 0 in the trilayer region. Finally, we observe similar $\mathbf{E}$-dependent nonlinear redshift of PL in two other samples which leads us to conclude that it is a generic feature of WS$_2$/WSe$_2$/WS$_2$ heterotrilayers (see Supplementary). Unlike the dipolar exciton in the bilayer region, $d\mathcal{E}/d E_{\mathrm{z}}$ (or the magnitude of dipole moment) in the trilayer region steadily increases from zero with increasing $|\mathbf{E}|$, for small $|\mathbf{E}|$. This behavior is consistent with that of a quadrupolar exciton, which also has a vanishing dipole moment at zero $\mathbf{E}$ (see Fig.~1a). 

To gain a qualitative understanding of the nonlinear redshift, we start by considering uncoupled top and bottom dipolar excitons with opposite dipole moments. Under an applied $\mathbf{E}$, the energy of the two dipolar excitons should shift in opposite directions, resulting in `X'-shaped dispersing branches (Fig.~2c). If we assume a finite hybridization of the two branches due to layer-hybridization of electrons, the intersecting `X'-like branches should turn into avoided crossing where the lower (higher) energy branch corresponds to the symmetric (antisymmetric) superposition of top and bottom dipolar excitons -- quadrupolar excitons~\cite{SlobodkinPRL2020}. In this case, the lower (higher) energy symmetric (antisymmetric) branch only redshifts (blueshifts), asymptotically merging with the dipolar branch (Fig.~2c). Thus, we conclude that the redshifting of PL in the trilayer region is consistent with a symmetric quadrupolar exciton. The antisymmetric quadrupolar branch is at higher energy and is expected to be much weaker in emission under non-resonant excitation due to relaxation to the symmetric branch. Furthermore, oscillator strength, which is characterized by electron-hole overlap, is drastically reduced for the antisymmetric quadrupolar exciton because of the presence of a node in the electronic wavefunction at the location of the hole in the WSe$_2$ layer (Fig.~1c). The combination of these two effects possibly renders the antisymmetric quadrupolar exciton optically dark in our experiments. Using the above picture, we can get a rough estimate of the strength of hybridization or tunnel coupling, $t_0$, between the two dipolar excitons by fitting the energy shift of the symmetric quadrupolar branch using a hyperbolic function, $\mathcal{E}(\mathbf{E}) = -\sqrt{t_0^2 + \alpha (e\mathbf{E}\cdot \mathbf{d_0})^2}$, which yields $t_0 \sim$ 35 meV and $d_0$ is the bare dipole moment.

While the above analysis is performed assuming a constant hybridization, $t_0$, of opposite dipolar excitons, we can obtain a more accurate picture by considering how the layer-hybridized electronic wavefunction evolves with $\mathbf{E}$. Under an out-of-plane electric field, we can assume that the hole distribution remains unchanged and hence the excitonic energy shift is primarily determined by the changes to the electronic wavefunction. To this end, we performed DFT simulations to calculate the electronic wavefunction of the symmetric state as a function of $\mathbf{E}$ (see Methods). Figure 2d-e shows the electronic charge distribution at two values of $\mathbf{E}$. As expected, with increasing $|\mathbf{E}|$, the electronic charge distribution becomes asymmetric about the hole in the middle WSe$_2$ layer,   resulting in an increased dipole moment. We calculate the energy shift for the symmetric quadrupolar exciton from this $\mathbf{E}$-dependent dipole moment as shown in Fig.~2f and find very good agreement with our experimental results. We also note that our DFT calculations neglect moir\'{e} potential-related effects, suggesting that the latter do not play an essential role in the formation of quadrupolar excitons. 




To further confirm that the observed behavior in the trilayer region is indeed due to tunnel coupling, we fabricated a sample with AB stacking order of the outer WS$_2$ layers, which should suppress layer hybridization due to spin-conserving tunneling. As shown in Fig.~2g, we observe dipolar exciton-like response under $\mathbf{E}$ with the PL displaying linear red and blueshift depending on the direction of $\mathbf{E}$, as expected from a lack of tunnel coupling.

Having established the existence of quadrupolar excitons in heterotrilayers, we study the implications of quadrupolar exciton wavefunction on light-matter coupling. A key quantity in determining the latter is the overlap of electron and hole wavefunction, which determines the radiative lifetime of the exciton. Owing to the tunneling of electrons through the WSe$_2$ barrier which hosts holes, the symmetric quadrupolar exciton has a larger electron-hole overlap compared to the dipolar exciton. As the electron-hole overlap in the symmetric quadrupolar exciton depends sensitively on $\mathbf{E}$, it can be tuned to control excitonic radiative lifetime and hence light-matter coupling. 

To test this hypothesis, we performed time-resolved PL lifetime measurement of the emission from bilayer and trilayer regions as a function of $\mathbf{E}$ (see Methods). Figure 3a  shows a measured PL lifetime of $\sim$0.4 ns in the trilayer region at $\mathbf{E}$ = 0, which is similar in the trilayer region of Device 2 (see Supplementary). The bilayer PL lifetime, on the other hand, is measured to be $\sim$ 1.2 ns. The longer PL lifetime of the bilayer emission is consistent with it being a dipolar exciton, which has reduced electron-hole overlap~\cite{RiveraNComm2015}. As shown in Fig.~3c and 3e, with increasing $|\mathbf{E}|$, PL lifetime of the heterotrilayer steadily increases to $\sim$0.7 ns. As the PL lifetime depends on both radiative and non-radiative lifetimes, an increase in PL lifetime could arise from a reduction in non-radiative processes or an increase in radiative lifetime, or a combination thereof. The presence of an out-of-plane electric field should have a minor affect on non-radiative lifetime. In fact, any unintentional doping under $\mathbf{E}$ will only decrease the non-radiative lifetime by carrier-induced relaxation. Thus, we attribute this increase in lifetime to reduction in electron-hole overlap of the quadrupolar exciton with a polarizing $|\mathbf{E}|$. The  trilayer region of Device 1 also shows an increased lifetime with $\mathbf{E}$ but by a smaller amount (see Supplementary). In contrast, Fig.~3d and 3f show that the bilayer region lifetime has negligible change for $V_{\mathrm{bg}} <$ 0. The decrease in PL lifetime of bilayer region for $V_{\mathrm{bg}} >$ 2V is consistent with the observed reduction in PL intensity in Fig.~2e and possibly arises from non-radiative relaxation due to  unintentional carrier doping. As the bilayer exciton is longer lived, it expected to be more sensitive to carrier-induced non-radiative channels.

As neutral quasiparticles, excitons in general interact weakly, but dipolar excitons can interact strongly via dipole-dipole interactions at sufficiently large excitonic densities~\cite{RiveraScience2016,liNNano2020dipolar,kremserNpj2D2020discrete}. Whereas interlayer dipolar excitons in heterobilayers interact repulsively, repulsive quadrupolar interactions between hybridized excitons in heterotrilayers are expected to be weaker. Exciton-exciton interactions can be studied by varying the steady-state excitonic density, $n_{\mathrm{ex}}$, which is proportional to $1/\bar{r}^{2}$, where $\bar{r}$ is the average inter-exciton distance. $n_{\mathrm{ex}}$ can be efficiently varied, for example, by changing the  excitation laser intensity, which is resonant with the intralayer excitonic resonance.


Figure 4 shows PL spectra of bilayer and trilayer regions for increasing excitation intensities of laser resonant with intralayer WSe$_2$ exciton. While the spectra of the bilayer excitons blueshift with increasing density (Fig.~4b), the spectra of the trilayer exciton (Fig.~4a) show negligible shift under the same excitation laser power. To estimate the interaction-induced energy shift for dipolar and quadrupolar excitons as a function of mean inter-exciton separation, $\bar{r}$, we assume a mean-field model and calculate the corresponding dipolar (quadrupolar) repulsive electrostatic energy, U$_{\mathrm{dd}}$ (U$_{\mathrm{qq}}$) (see Supplementary). As shown in Fig.~4c, for a 5 meV blueshift of dipolar exciton, the quadrupolar exciton shifts by 0.16 meV for the same exciton density. Considering the difference in lifetimes of dipolar and quadrupolar excitons, $\tau_{d} / \tau_{q} \sim 3$, the steady-state density of quadrupolar excitons is expected to be $\sim$ 3 times smaller than that of dipolar excitons. Thus, $\bar{r}_{\mathrm{q}}$ is $\sim$ $\sqrt{3} \bar{r}_{\mathrm{d}}$ for a fixed density of dipolar excitons. As a result, the quadrupolar blueshift at the same excitation intensity which causes a 5 meV blueshift in dipolar excitons reduces further to $\sim$ 0.01 meV. This analysis is consistent with the observed absence of blueshift in heterotrilayer excitons with increasing excitation/emission intensity. Finally, as shown in Fig.~4d the bilayer region PL exhibits a sublinear power-dependence and saturation behavior with excitation power while the trilayer region PL varies linearly and does not saturate up to the highest incident power. This behavior is in agreement with the longer PL lifetime of the bilayer exciton compared to that of the trilayer exciton.

In conclusion, we demonstrated the observation of an unconventional exciton in a vdW heterotrilayer with an out-of-plane quadrupolar moment. These quadrupolar excitons are formed as symmetric coherent superposition of oppositely oriented dipolar excitons. Due to their large polarizability, an external electric field can modify the wavefunction of the quadrupolar exciton wavefunction and change its dipole moment, which can be exploited for tunable light-matter interactions. In addition, excitonic wavefunction engineering in vdW heterotrilayers demonstrated here can serve as a versatile tool to tune many-exciton interactions and study strongly correlated phases of bosons and quantum phase transitions between them.

\textit{Note added}: Recently, we became aware of a manuscript from Stanford group reporting similar results in vdW heterotrilayers~\cite{yuArXiv2022}.

\justify
{\bf Acknowledgments} We thank Hayk Harutyunyan for help with lifetime measurements. This work was supported by the EFRI program-grant (\# EFMA-1741691 for A. S.) and NSF DMR award (\# 1905809 for A. S.). The theoretical work was supported by the European Research Council (ERC-2015-AdG694097), cluster of Excellence AIM, SFB925 and Grupos Consolidados (IT1249-19). We acknowledge support by the Max Planck Institute-New York City Center for Non-Equilibrium Quantum Phenomena. The Flatiron Institute is a division of the Simons Foundation. J.Z. acknowledges funding received from the European Union Horizon 2020 research and innovation program under Marie Sklodowska-Curie Grant Agreement 886291 (PeSD-NeSL). K.W. and T.T. acknowledge support from the Elemental Strategy Initiative conducted by the MEXT, Japan (Grant Number JPMXP0112101001) and  JSPS KAKENHI (Grant Numbers 19H05790, 20H00354 and 21H05233).
\justify
{\bf Author contributions}
A. S., W. L., L. D., Z. H. conceived the project. K. W., T. T. provided the hBN crystal and S. L. and J. H. provided the WSe$_2$ crystals. W. L., Z. H., L. D. prepared the samples. W. L., Z. H., L. D. carried out the measurements. J. Z. conducted the DFT calculations. A. S., A. R. supervised the project. All authors were involved in analysis of the experimental data and contributed extensively.




\begin{thebibliography}{10}
\expandafter\ifx\csname url\endcsname\relax
  \def\url#1{\texttt{#1}}\fi
\expandafter\ifx\csname urlprefix\endcsname\relax\def\urlprefix{URL }\fi
\providecommand{\bibinfo}[2]{#2}
\providecommand{\eprint}[2][]{\url{#2}}

\bibitem{wangRMP2018colloquium}
\bibinfo{author}{Wang, G.} \emph{et~al.}
\newblock \bibinfo{title}{Colloquium: Excitons in atomically thin transition
  metal dichalcogenides}.
\newblock \emph{\bibinfo{journal}{Reviews of Modern Physics}}
  \textbf{\bibinfo{volume}{90}}, \bibinfo{pages}{021001}
  (\bibinfo{year}{2018}).

\bibitem{MakNMat2013}
\bibinfo{author}{Mak, K.~F.} \emph{et~al.}
\newblock \bibinfo{title}{Tightly bound trions in monolayer MoS$_2$}.
\newblock \emph{\bibinfo{journal}{Nat Mater}} \textbf{\bibinfo{volume}{12}},
  \bibinfo{pages}{207--211} (\bibinfo{year}{2013}).
\newblock

\bibitem{HePRL2014}
\bibinfo{author}{He, K.} \emph{et~al.}
\newblock \bibinfo{title}{Tightly bound excitons in monolayer
  ${\mathrm{WSe}}_{2}$}.
\newblock \emph{\bibinfo{journal}{Phys. Rev. Lett.}}
  \textbf{\bibinfo{volume}{113}}, \bibinfo{pages}{026803}
  (\bibinfo{year}{2014}).
\newblock

\bibitem{RiveraScience2016}
\bibinfo{author}{Rivera, P.} \emph{et~al.}
\newblock \bibinfo{title}{Valley-polarized exciton dynamics in a 2d
  semiconductor heterostructure}.
\newblock \emph{\bibinfo{journal}{Science}} \textbf{\bibinfo{volume}{351}},
  \bibinfo{pages}{688--691} (\bibinfo{year}{2016}).
\newblock 

\bibitem{liNNano2021optical}
\bibinfo{author}{Li, W.}, \bibinfo{author}{Lu, X.}, \bibinfo{author}{Wu, J.} \&
  \bibinfo{author}{Srivastava, A.}
\newblock \bibinfo{title}{Optical control of the valley zeeman effect through
  many-exciton interactions}.
\newblock \emph{\bibinfo{journal}{Nature Nanotechnology}}
  \textbf{\bibinfo{volume}{16}}, \bibinfo{pages}{148--152}
  (\bibinfo{year}{2021}).


\bibitem{kremserNpj2D2020discrete}
\bibinfo{author}{Kremser, M.} \emph{et~al.}
\newblock \bibinfo{title}{Discrete interactions between a few interlayer
  excitons trapped at a MoSe$_2$--WSe$_2$ heterointerface}.
\newblock \emph{\bibinfo{journal}{npj 2D Materials and Applications}}
  \textbf{\bibinfo{volume}{4}}, \bibinfo{pages}{1--6} (\bibinfo{year}{2020}).

\bibitem{sunNPhot2022excitonic}
\bibinfo{author}{Sun, Z.} \emph{et~al.}
\newblock \bibinfo{title}{Excitonic transport driven by repulsive dipolar interaction in a van der Waals heterostructure}.
\newblock \emph{\bibinfo{journal}{Nature Photonics}}
  \textbf{\bibinfo{volume}{16}}, \bibinfo{pages}{79--85}
  (\bibinfo{year}{2022}).

\bibitem{FangPNAS2014}
\bibinfo{author}{Fang, H.} \emph{et~al.}
\newblock \bibinfo{title}{Strong interlayer coupling in van der Waals
  heterostructures built from single-layer chalcogenides}.
\newblock \emph{\bibinfo{journal}{Proceedings of the National Academy of
  Sciences}} \textbf{\bibinfo{volume}{111}}, \bibinfo{pages}{6198--6202}
  (\bibinfo{year}{2014}).
\newblock

\bibitem{RiveraNComm2015}
\bibinfo{author}{Rivera, P.} \emph{et~al.}
\newblock \bibinfo{title}{Observation of long-lived interlayer excitons in
  monolayer MoSe$_2$--WSe$_2$ heterostructures}.
\newblock \emph{\bibinfo{journal}{Nature Communications}}
  \textbf{\bibinfo{volume}{6}}, \bibinfo{pages}{6242 EP --}
  (\bibinfo{year}{2015}).
\newblock

\bibitem{WangNanoLett2018}
\bibinfo{author}{Wang, Z.}, \bibinfo{author}{Chiu, Y.-H.},
  \bibinfo{author}{Honz, K.}, \bibinfo{author}{Mak, K.~F.} \&
  \bibinfo{author}{Shan, J.}
\newblock \bibinfo{title}{Electrical tuning of interlayer exciton gases in WSe$_2$ bilayers}.
\newblock \emph{\bibinfo{journal}{Nano Letters}} \textbf{\bibinfo{volume}{18}},
  \bibinfo{pages}{137--143} (\bibinfo{year}{2018}).
\newblock

\bibitem{CiarrocchiNP2019}
\bibinfo{author}{Ciarrocchi, A.} \emph{et~al.}
\newblock \bibinfo{title}{Polarization switching and electrical control of
  interlayer excitons in two-dimensional van der Waals heterostructures.}
\newblock \emph{\bibinfo{journal}{Nature photonics}}
  \textbf{\bibinfo{volume}{13}}, \bibinfo{pages}{131--136}
  (\bibinfo{year}{2019}).
\newblock

\bibitem{BaranowskiNanoLett2017}
\bibinfo{author}{Baranowski, M.} \emph{et~al.}
\newblock \bibinfo{title}{Probing the interlayer exciton physics in a
  MoS$_2$/MoSe$_2$/MoS$_2$ van der Waals heterostructure}.
\newblock \emph{\bibinfo{journal}{Nano Letters}} \textbf{\bibinfo{volume}{17}},
  \bibinfo{pages}{6360--6365} (\bibinfo{year}{2017}).
\newblock

\bibitem{SlobodkinPRL2020}
\bibinfo{author}{Slobodkin, Y.} \emph{et~al.}
\newblock \bibinfo{title}{Quantum phase transitions of trilayer excitons in
  atomically thin heterostructures}.
\newblock \emph{\bibinfo{journal}{Physical Review Letters}}
  \textbf{\bibinfo{volume}{125}}, \bibinfo{pages}{255301}
  (\bibinfo{year}{2020}).

\bibitem{SammonPRB2019}
\bibinfo{author}{Sammon, M.} \& \bibinfo{author}{Shklovskii, B.~I.}
\newblock \bibinfo{title}{Attraction of indirect excitons in van der Waals
  heterostructures with three semiconducting layers}.
\newblock \emph{\bibinfo{journal}{Physical Review B}}
  \textbf{\bibinfo{volume}{99}}, \bibinfo{pages}{165403}
  (\bibinfo{year}{2019}).
  
  \bibitem{astrakharchikPRB2021quantum}
\bibinfo{author}{Astrakharchik, G.}, \bibinfo{author}{Kurbakov, I.},
  \bibinfo{author}{Sychev, D.}, \bibinfo{author}{Fedorov, A.} \&
  \bibinfo{author}{Lozovik, Y.~E.}
\newblock \bibinfo{title}{Quantum phase transition of a two-dimensional
  quadrupolar system}.
\newblock \emph{\bibinfo{journal}{Physical Review B}}
  \textbf{\bibinfo{volume}{103}}, \bibinfo{pages}{L140101}
  (\bibinfo{year}{2021}).

\bibitem{yongNMat2019valley}
\bibinfo{author}{Yong, C.-K.} \emph{et~al.}
\newblock \bibinfo{title}{Valley-dependent exciton fine structure and
  autler--townes doublets from berry phases in monolayer MoSe$_2$}.
\newblock \emph{\bibinfo{journal}{Nature materials}}
  \textbf{\bibinfo{volume}{18}}, \bibinfo{pages}{1065--1070}
  (\bibinfo{year}{2019}).

\bibitem{guNComm2021enhanced}
\bibinfo{author}{Gu, J.} \emph{et~al.}
\newblock \bibinfo{title}{Enhanced nonlinear interaction of polaritons via
  excitonic rydberg states in monolayer WSe$_2$}.
\newblock \emph{\bibinfo{journal}{Nature communications}}
  \textbf{\bibinfo{volume}{12}}, \bibinfo{pages}{1--7} (\bibinfo{year}{2021}).

\bibitem{ZhangNature2021}
\bibinfo{author}{Zhang, L.} \emph{et~al.}
\newblock \bibinfo{title}{Van der waals heterostructure polaritons with
  moir{\'e}-induced nonlinearity}.
\newblock \emph{\bibinfo{journal}{Nature}} \textbf{\bibinfo{volume}{591}},
  \bibinfo{pages}{61--65} (\bibinfo{year}{2021}).

\bibitem{TanPRX2020}
\bibinfo{author}{Tan, L.~B.} \emph{et~al.}
\newblock \bibinfo{title}{Interacting polaron-polaritons}.
\newblock \emph{\bibinfo{journal}{Physical Review X}}
  \textbf{\bibinfo{volume}{10}}, \bibinfo{pages}{021011}
  (\bibinfo{year}{2020}).

\bibitem{dagvadorjPRB2021first}
\bibinfo{author}{Dagvadorj, G.}, \bibinfo{author}{Kulczykowski, M.},
  \bibinfo{author}{Szyma{\'n}ska, M.~H.} \& \bibinfo{author}{Matuszewski, M.}
\newblock \bibinfo{title}{First-order dissipative phase transition in an
  exciton-polariton condensate}.
\newblock \emph{\bibinfo{journal}{Physical Review B}}
  \textbf{\bibinfo{volume}{104}}, \bibinfo{pages}{165301}
  (\bibinfo{year}{2021}).

\bibitem{AlexeevNature2019}
\bibinfo{author}{Alexeev, E.~M.} \emph{et~al.}
\newblock \bibinfo{title}{Resonantly hybridized excitons in moir{\'e}
  superlattices in van der Waals heterostructures}.
\newblock \emph{\bibinfo{journal}{Nature}} \textbf{\bibinfo{volume}{567}},
  \bibinfo{pages}{81--86} (\bibinfo{year}{2019}).

\bibitem{ShimazakiNature2020}
\bibinfo{author}{Shimazaki, Y.} \emph{et~al.}
\newblock \bibinfo{title}{Strongly correlated electrons and hybrid excitons in
  a moir{\'e} heterostructure}.
\newblock \emph{\bibinfo{journal}{Nature}} \textbf{\bibinfo{volume}{580}},
  \bibinfo{pages}{472--477} (\bibinfo{year}{2020}).

\bibitem{HsuSciAdv2019tailor}
\bibinfo{author}{Hsu, W.-T.} \emph{et~al.}
\newblock \bibinfo{title}{Tailoring excitonic states of van der waals bilayers
  through stacking configuration, band alignment, and valley spin}.
\newblock \emph{\bibinfo{journal}{Science advances}}
  \textbf{\bibinfo{volume}{5}}, \bibinfo{pages}{eaax7407}
  (\bibinfo{year}{2019}).

\bibitem{XiaoPRL2012}
\bibinfo{author}{Xiao, D.}, \bibinfo{author}{Liu, G.-B.},
  \bibinfo{author}{Feng, W.}, \bibinfo{author}{Xu, X.} \& \bibinfo{author}{Yao,
  W.}
\newblock \bibinfo{title}{Coupled spin and valley physics in monolayers of
  ${\mathrm{MoS}}_{2}$ and other group-vi dichalcogenides}.
\newblock \emph{\bibinfo{journal}{Phys. Rev. Lett.}}
  \textbf{\bibinfo{volume}{108}}, \bibinfo{pages}{196802}
  (\bibinfo{year}{2012}).
\newblock

\bibitem{yuOptica2020observation}
\bibinfo{author}{Yu, J.} \emph{et~al.}
\newblock \bibinfo{title}{Observation of double indirect interlayer exciton in
  WSe$_2$/WS$_2$ heterostructure}.
\newblock \emph{\bibinfo{journal}{Optics express}}
  \textbf{\bibinfo{volume}{28}}, \bibinfo{pages}{13260--13268}
  (\bibinfo{year}{2020}).

\bibitem{ParadisanosNComm2021}
\bibinfo{author}{Paradisanos, I.} \emph{et~al.}
\newblock \bibinfo{title}{Efficient phonon cascades in wse2 monolayers}.
\newblock \emph{\bibinfo{journal}{Nature communications}}
  \textbf{\bibinfo{volume}{12}}, \bibinfo{pages}{1--7} (\bibinfo{year}{2021}).

\bibitem{JaureguiScience2019}
\bibinfo{author}{Jauregui, L.~A.} \emph{et~al.}
\newblock \bibinfo{title}{Electrical control of interlayer exciton dynamics in
  atomically thin heterostructures}.
\newblock \emph{\bibinfo{journal}{Science}} \textbf{\bibinfo{volume}{366}},
  \bibinfo{pages}{870--875} (\bibinfo{year}{2019}).

\bibitem{liNNano2020dipolar}
\bibinfo{author}{Li, W.}, \bibinfo{author}{Lu, X.}, \bibinfo{author}{Dubey,
  S.}, \bibinfo{author}{Devenica, L.} \& \bibinfo{author}{Srivastava, A.}
\newblock \bibinfo{title}{Dipolar interactions between localized interlayer
  excitons in van der waals heterostructures}.
\newblock \emph{\bibinfo{journal}{Nature Materials}}
  \textbf{\bibinfo{volume}{19}}, \bibinfo{pages}{624--629}
  (\bibinfo{year}{2020}).

\bibitem{yuArXiv2022}
\bibinfo{author}{Yu.~L. et~al.,}
\newblock \bibinfo{title}{{to appear}}.
\newblock \emph{\bibinfo{journal}{ArXiv}}  (\bibinfo{year}{2022}).

\bibitem{zomer2014fast}
\bibinfo{author}{Zomer, P.}, \bibinfo{author}{Guimar{\~a}es, M.},
  \bibinfo{author}{Brant, J.}, \bibinfo{author}{Tombros, N.} \&
  \bibinfo{author}{Van~Wees, B.}
\newblock \bibinfo{title}{Fast pick up technique for high quality
  heterostructures of bilayer graphene and hexagonal boron nitride}.
\newblock \emph{\bibinfo{journal}{Applied Physics Letters}}
  \textbf{\bibinfo{volume}{105}}, \bibinfo{pages}{013101}
  (\bibinfo{year}{2014}).

\bibitem{kim2016van}
\bibinfo{author}{Kim, K.} \emph{et~al.}
\newblock \bibinfo{title}{van der waals heterostructures with high accuracy
  rotational alignment}.
\newblock \emph{\bibinfo{journal}{Nano letters}} \textbf{\bibinfo{volume}{16}},
  \bibinfo{pages}{1989--1995} (\bibinfo{year}{2016}).

\bibitem{KressePRB1996}
\bibinfo{author}{Kresse, G.} \& \bibinfo{author}{Furthm{\"u}ller, J.}
\newblock \bibinfo{title}{Efficient iterative schemes for ab initio
  total-energy calculations using a plane-wave basis set}.
\newblock \emph{\bibinfo{journal}{Physical review B}}
  \textbf{\bibinfo{volume}{54}}, \bibinfo{pages}{11169} (\bibinfo{year}{1996}).

\bibitem{PerdewPRL1996}
\bibinfo{author}{Perdew, J.~P.}, \bibinfo{author}{Burke, K.} \&
  \bibinfo{author}{Ernzerhof, M.}
\newblock \bibinfo{title}{Generalized gradient approximation made simple}.
\newblock \emph{\bibinfo{journal}{Physical review letters}}
  \textbf{\bibinfo{volume}{77}}, \bibinfo{pages}{3865} (\bibinfo{year}{1996}).

\bibitem{KlimevsPRB2011}
\bibinfo{author}{Klime{\v{s}}, J.}, \bibinfo{author}{Bowler, D.~R.} \&
  \bibinfo{author}{Michaelides, A.}
\newblock \bibinfo{title}{Van der waals density functionals applied to solids}.
\newblock \emph{\bibinfo{journal}{Physical Review B}}
  \textbf{\bibinfo{volume}{83}}, \bibinfo{pages}{195131}
  (\bibinfo{year}{2011}).

\end{thebibliography}




\clearpage

\justify

\textbf{Methods}
\\
\textbf{Device fabrication} \\
Layer-by-layer dry transfer using a polycarbonate (PC) stamp is used to fabricate the dual-gated transition metal dichalcogenide heterotrilayer devices~\cite{zomer2014fast}. WSe$_2$ and WS$_2$ monolayers, few-layer graphene and thick hBN are mechanically exfoliated from bulk crystals onto 300 nm SiO$_2$/Si substrates. The thickness of the flakes is determined by optical contrast, and we estimate the hBN thickness to be 20-40 nm. For the sample assembly, we use a PC stamp to pick up the bottom hBN and the bottom few-layer graphene, which is then placed onto a 300 nm SiO$_2$/Si substrate with pre-patterned electrodes (5 nm Cr/85 nm Au) at 170 $^\circ$C. Afterwards, the substrate with bottom hBN and graphene is annealed in 5\%H$_2$/95\%N$_2$ at 350 $^\circ$C for 3 hours to remove the PC residue. The few-layer graphene top gate, top hBN flake, top WS$_2$ monolayer, WSe$_2$ monolayer, bottom WS$_2$ monolayer, the few-layer-graphene contact is picked up with a PC stamp and then placed onto the back gate at 170 $^\circ$C. We use the tear-and-stack method~\cite{kim2016van} to pick up half of a WS$_2$ flake, align with the WSe$_2$ layer within 1$^\circ$ uncertainty and pick up the other half of the WS$_2$ flake without rotation (AA stacking, Device 1, 2 in the main text, Device 4 in the supplementary) and with 60$^\circ$ rotation (AB stacking, Device 3 in the main text) relative to the top WS$_2$ in order to protect/break the mirror symmetry in AA/AB samples. All data in the main text are from Device 1 with the exception of Fig.~3e, f which are from Device 2, and Fig.~2g from Device 3. 

\justify
\textbf{Optical measurements and electrostatic gating} \\
Photoluminescence and reflection contrast sprectrosocopy measurements is performed inside a cryostat (AttoDry 800, 6K). We use a piezoelectric controller (Attocube systems) to position the sample. A mode-hop-free tunable continuous-wave Ti:Sapphire laser (MSquared Lasers) with a wavelength resolution of 0.1 pm is used as the excitation laser for the photoluminescence measurements. A halogen lamp (Thorlabs SLS01L) serves as the white light source for the reflectance contrast measurements. An achromatic objective (NA = 0.42) is used to focus the laser and white light beams to a spot size of $\sim$1 $\mu$m, which is then collected through the same objective (focal length 500 mm) and directed to a high-resolution spectrometer (Princetron Instrument HR-500) which disperses the light by a 300 grooves per mm grating (blazed at 750 nm). A charge coupled device (Princeton Instrument PIXIS-400 CCD) is used as a detector. We control the circular polarization of the incident laser by a polarizer and a $\lambda$/4 waveplate.

Through Keithley 2400 source meters, we apply voltages to the graphene top gate and graphene bottom gate to tune the charge density or apply an electric field to the sample. Negligible doping is measured with antisymmetric gating voltages because the top and bottom gates are nearly symmetric with 20-40nm hBN gate dielectrics. 

\justify
\textbf{DFT calculations} \\
The ab initio calculations were performed within the Vienna Ab initio Simulation Package (VASP)\cite{KressePRB1996} using a projector-augmented wave (PAW) pseudopotential in conjunction with the Perdew–Burke–Ernzerhof (PBE)\cite{PerdewPRL1996} functionals and a plane-wave basis set with an energy cutoff at 400 eV.  The unit cells are chosen to consist WS$_2$/WSe$_2$/WS$_2$ trilayers with a lattice constant of 3.154 . A vacuum region of 20 Å is applied to avoid artificial interaction between the periodic images along the vertical direction. The first Brillouin zone of the heterostructure was sampled using a 15$\times$15$\times$1 k-point grids. The structures at ground states were fully relaxed until the force on each atom was $<$ 0.01 eV Å$^{-1}$. The van der Waals interactions were included using the opt88 functional\cite{KlimevsPRB2011}. For the structures with modified interlayer hybridizations, the interlayer distances were artificially modulated to get insight into the influence various stackings. Spin-orbital couplings are included in the calculations of electronic structures.

\justify
\textbf{Data availability}
\\
The data that support the plots within this paper and other findings of this study are available from the corresponding author upon reasonable request.

\clearpage

\begin{figure}
\includegraphics[scale=0.9]{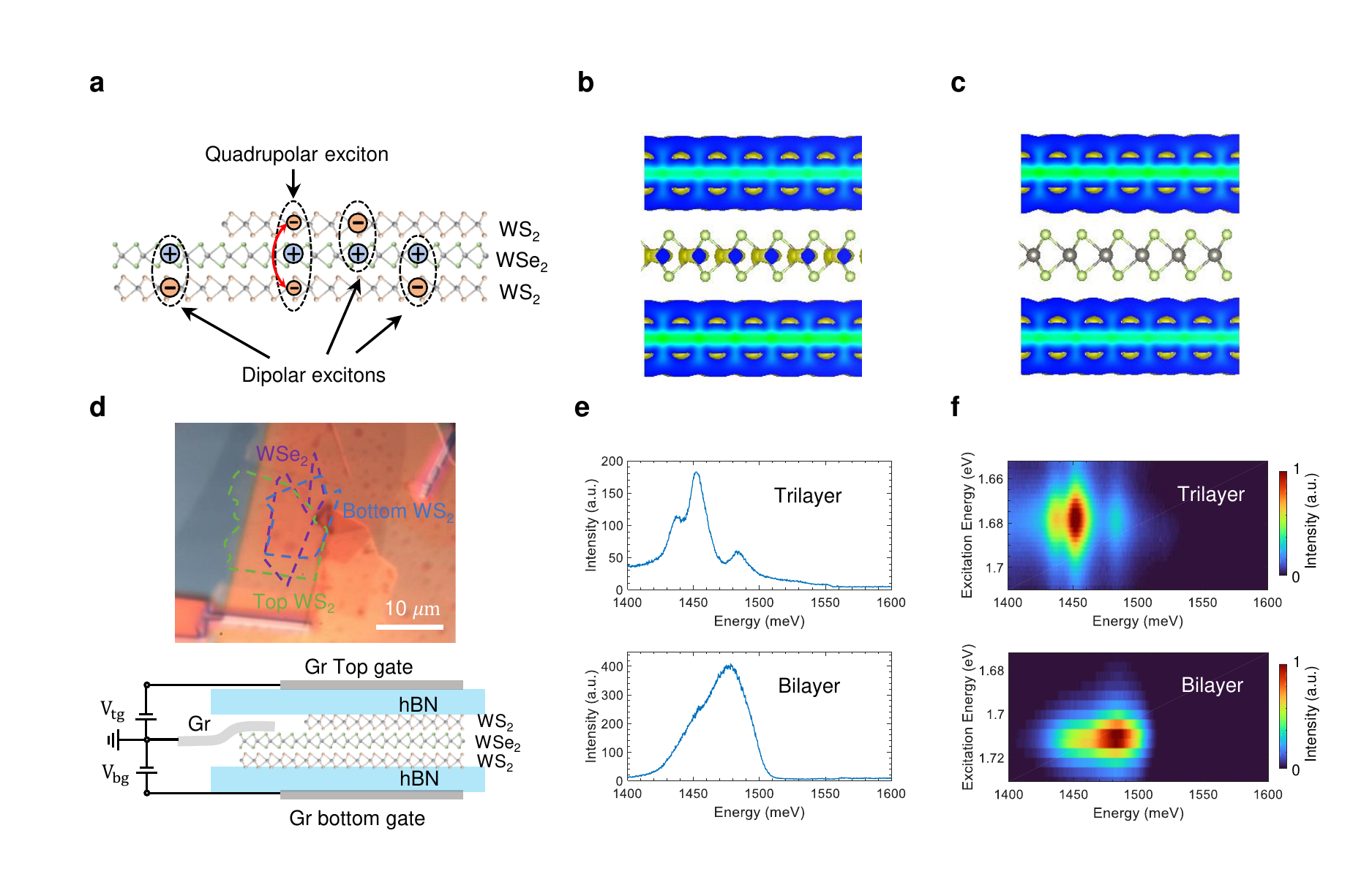}
\justify
\noindent {\bf Figure 1: Quadrupolar and dipolar excitons in a TMD heterostructure.}
{\bf a.}
Schematics of exciton species in bilayer and trilayer heterostructures. The bilayer region hosts one species of dipolar excitons, while the trilayer region can host two antiparallel dipolar excitons, as well as the quadrupolar exciton, which is a hybridized state of the two dipolar excitons.
{\bf b.}
Electron wavefunction distribution in symmetric quadrupolar excitons, when delocalized across the two WS$_2$ layers with a finite weight in the middle WSe$_2$ layers.
{\bf c.}
Electron wavefunction distribution in antisymmetric quadrupolar excitons, when delocalized across the two WS$_2$ layers with a vanishing weight in the middle WSe$_2$ layers.
{\bf d.}
Sample picture and gate configurations. The dual gates allow independent application of doping and electric field to the sample.
{\bf e.}
Representative photoluminescence (PL) spectra from the trilayer (top) and bilayer sample regions (bottom). The trilayer shows a three-peak structure, as opposed to the bilayer emission. Excitation was 40$\mu$W of 1.68 eV (1.71 eV) laser for the trilayer (bilayer). The bilayer peak is bluer compared to the strongest trilayer peak.
{\bf f.}
Photoluminescence excitation spectroscopy of the trilayer and bilayer region at $\mathbf{E}$ = 0. Both regions show one resonance, which is 1.68 eV (1.71 eV) for the trilayer (bilayer).

\end{figure}

\clearpage

\begin{figure}
\includegraphics[scale=0.9]{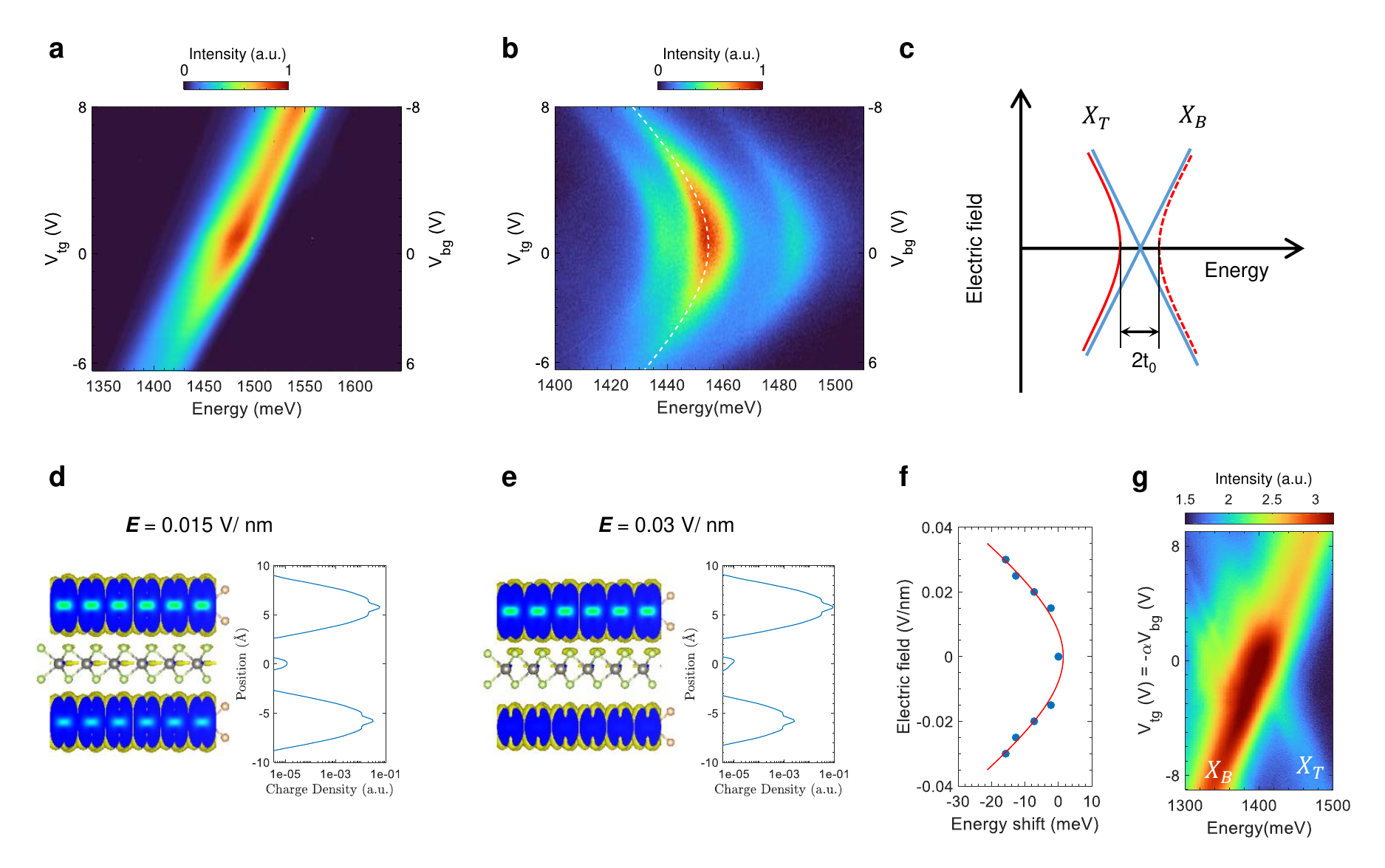}
 
\end{figure}
\noindent {\bf Figure 2: Electrical field tunable dipole hybridization in quadrupolar excitons.} 
{\bf a.}
Electric field ($\mathbf{E}$) dependent PL of bilayer excitons. The peak blueshifts or redshifts depending on the direction of the out-of-plane electric field. 
{\bf b.}
$\mathbf{E}$-dependent PL of trilayer excitons. The dashed white line is a hyperbolic fit of the main peak. The three-peak structure only redshifts with electric field. 
{\bf c.}
Hybridization of the top and bottom dipoles. $X_B$ ($X_T$) is the bottom (top) dipole and $t_0$ is the tunnelling strength at zero electric field. The solid red line is the lower energy symmetric branch, and the dashed red line is the higher energy antisymmetric branch. 
{\bf e, d.}
DFT calculation of the electron charge density across the layers at $\mathbf{E}=0.015 $V/nm (e) and $\mathbf{E}=0.03 $V/nm (d). When $\mathbf{E}$ is increased, the electron density is shifted from the bottom layer to the top layer, and the electron density in the middle layer is reduced. 
{\bf f.}
The solid blue dots is the energy shifts of the quadrupolar exciton as a function of $\mathbf{E}$ from DFT calculations. The red line shows a hyperbolic fit. 
{\bf g.}
$\mathbf{E}$-dependent PL of an AB stacked trilayer (Device 3). The top and bottom dipoles do not hybridize. $\alpha=1.289$ is for applying $\mathbf{E}$ without doping. The excitation power is 40$\mu$W for {\bf(a)} and {\bf(b)}, and the excitation energies are 1.71 eV and 1.68 eV, respectively. The excitation power is 5$\mu$W for {\bf(g)} with energy 1.685 eV. 

\clearpage

\begin{figure}
\includegraphics[scale=0.9]{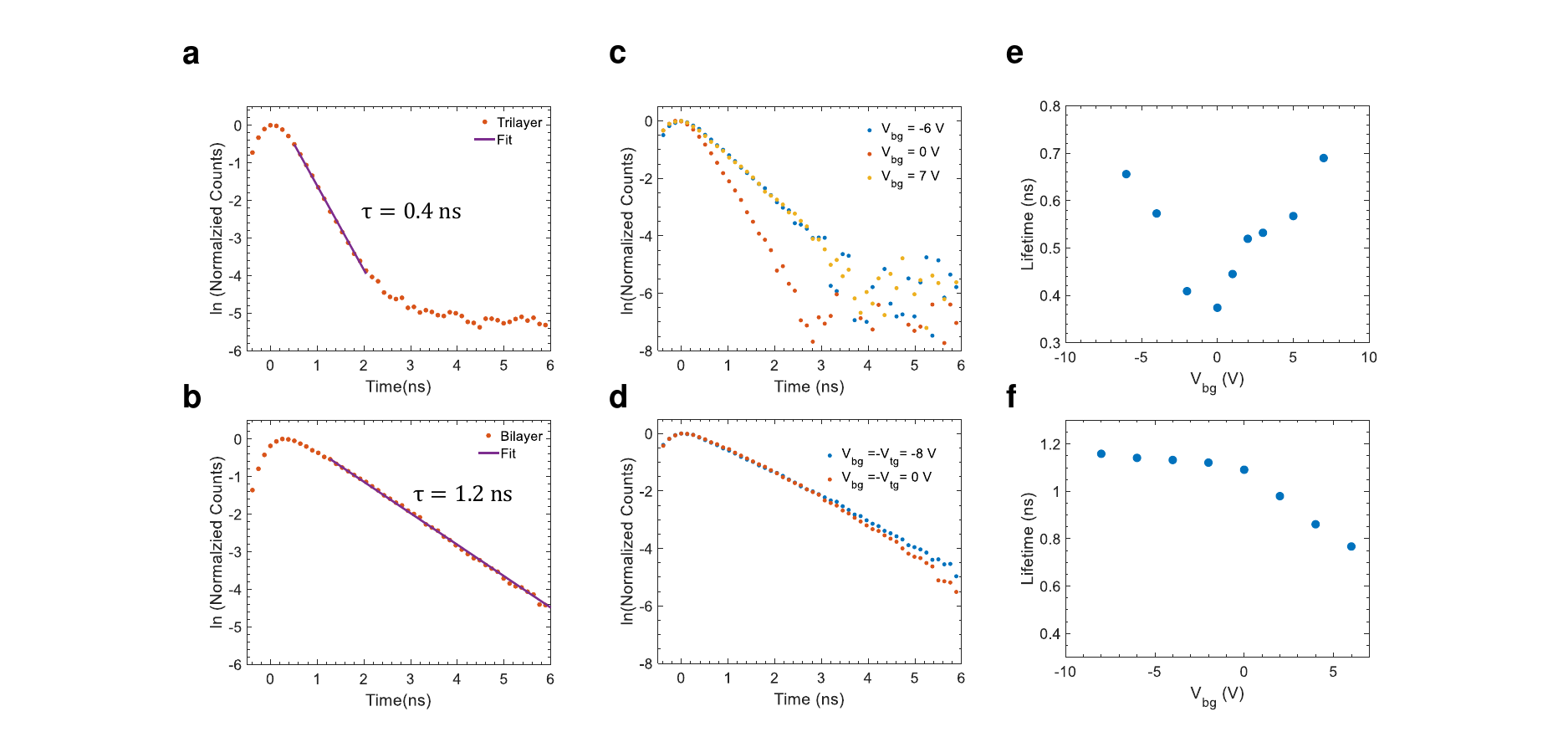}
\end{figure}

\noindent {\bf Figure 3: Electrical control of quadrupolar and dipolar exciton lifetimes.}
{\bf a.}
Lifetime measurement of the trilayer exciton at zero electric field ($\mathbf{E}$. Exponential fit yields a lifetime of 0.4 ns.
{\bf b.}
Lifetime measurement of the bilayer exciton at $\mathbf{E}$ = 0. Exponential fit yields a lifetime of 1.2 ns.
{\bf c.}
Time-resolved PL of the trilayer exciton with a positive, negative and zero V$_{\mathrm{bg}}$.
{\bf d.}
Time-resolved PL of the bilayer exciton as a function with $\mathbf{E}$.
{\bf e.}
Fitted lifetime of the trilayer exciton as a function of V$_{\mathrm{bg}}$. The trilayer exciton lifetime increases with $\mathbf{E}$ in either direction.
{\bf f.}
Fitted lifetime of the bilayer exciton as a function of V$_{\mathrm{bg}}$. The bilayer lifetime has negligible $\mathbf{E}$ dependence.
\clearpage

\begin{figure}
\includegraphics[scale=0.9]{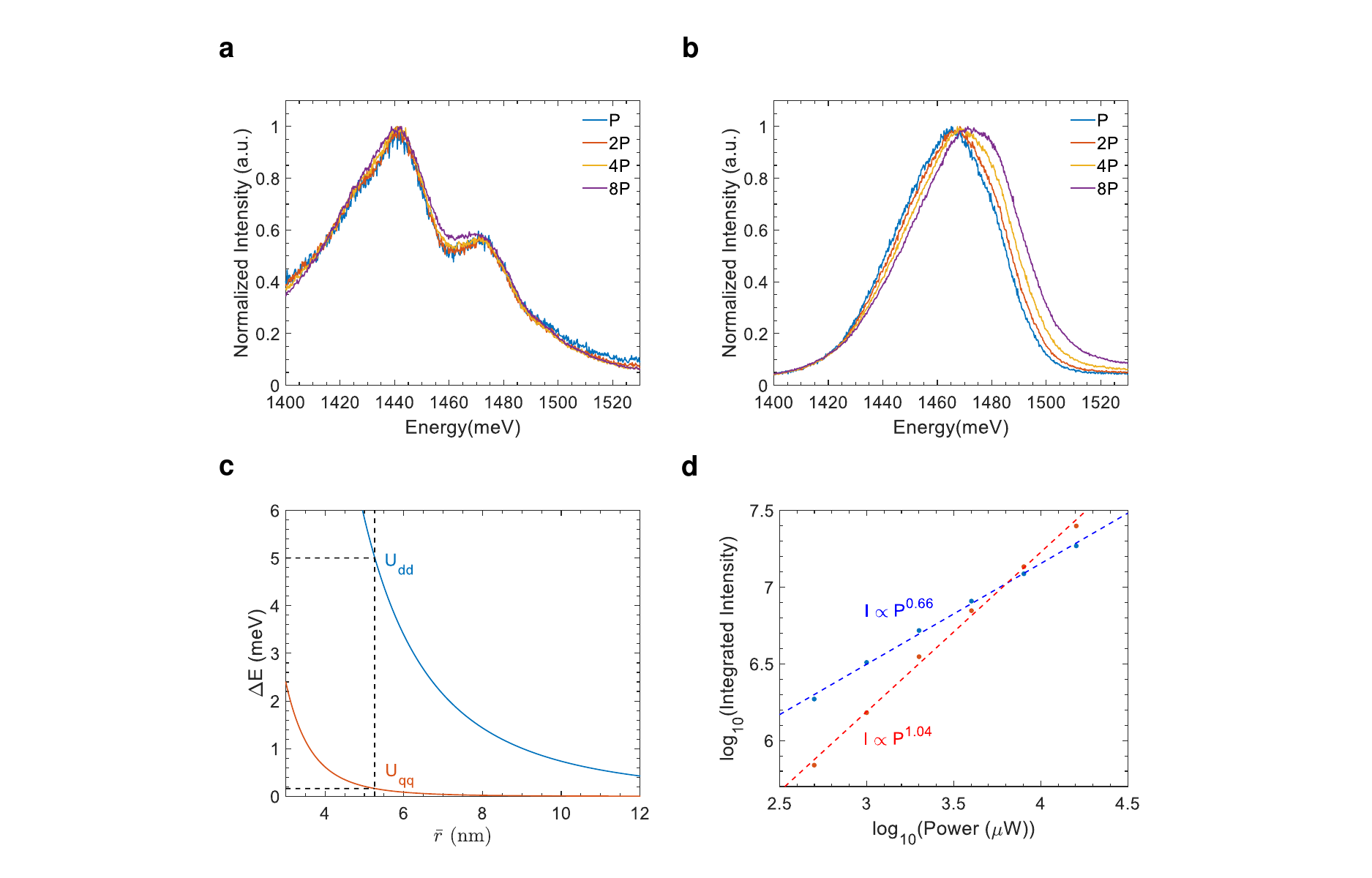}
\end{figure}
\noindent {\bf Figure 4: Density dependent energy shifts of quadrupolar and dipolar excitons.}
{\bf a.} Normalized spectra from power dependence measurements of trilayer emission, P = 0.5 mW. Peaks show no clear shift. {\bf b.} Normalized spectra for bilayer power dependence. The emission blueshifts with power. {\bf c.} Calculated dipolar (U$_\mathrm{dd}$) and quadrupolar (U$_\mathrm{qq}$) energy shifts as a function of average inter-exciton distance, $\bar{r}$. {\bf d.} Log-log plot of integrated intensity of the trilayer (red solid dots) and bilayer (blue solid dots) emission versus power. The power law fit for the trilayer (red dashed line) and bilayer (blue dashed line) is linear and sublinear, respectively . 
\clearpage

\end{document}